# Review of Design of Speech Recognition and Text Analytics based Digital Banking Customer Interface and Future Directions of Technology Adoption

Amal K Saha

Faculty of Physical Sciences, SGT University, Gurgaon, India ; previously worked as senior architect in information technology division of multinational banks

e-mail:  amal.k.saha@gmail.com

**Abstract.**  Banking is one of the most significant adopters of cutting-edge information technologies. Since its modern era beginning in the form of paper based accounting maintained in the branch, adoption of computerized system made it possible to centralize the processing in data centre and improve customer experience by making a more available and efficient system. The latest twist in this evolution is adoption of natural language processing and speech recognition in the user interface between the human and the system and use of machine learning and advanced analytics, in general, for backend processing as well. The paper reviews the progress of technology adoption in the field and comments on the maturity level of solutions involving less studied or low-resource languages like Hindi and also other Indian, regional languages. Furthermore, it also provides an analysis from a prototype built by us. The future directions of this area are also highlighted.

# 1. Introduction

Banking has been consistently employing cutting-edge information technologies. It started with paper based accounting maintained at branch level, and centralization happened with the adoption of computerized system, with the processing taking place in data centre. It improved customer experience significantly [2] by making a more available and efficient system. The customer really appreciated anytime access to banking functionalities through the internet. But the pace of innovation in technology in backend processing and also user interface exceeded the expectation of many customers and the trend continued. The support for mobile devices of various form factors and computers of standard sizes defined new customer experience. This revolutionized the banking functions.

The latest twist in this is interest in adopting natural language processing and speech recognition as the interface between the human and the system and use of advanced analytics and artificial intelligence (AI), in general, for backend processing as well. The new interface mentioned here goes by the name of conversational user interface [20]. In backend, the daily operations like beginning-of-day, end-of-day and near real-time processing had already been introduced more than three decades back. What is new is the ability of the system to learn from data. Normal solutions are programmed with fixed rules - if the condition is satisfied, do this, else do that. They are called rule-based solutions. The other types of solutions known as cognitive solutions, on the other hand, are trained and not programmed as in rule-based systems — they learn with interactions and new pieces of information. This supervised learning often involves intensive training of the system by human domain expert during training phase. The data generated by human expert during training helps the system recognize the solution patterns and handle situation involving unknown data. Once this training is done properly, the system scales the highest level of human expertise with high degree of precision and repeatability.

That journey of information technology in banking did not stop there. It started mimicking the function of human interaction by supporting speech recognition and text analysis. Collectively, these features of human-like interface and the intelligent backend processing driven by learning from data, are what folks in the banking industry call cognitive banking [1]. Cognitive computing [10] is a computerized model that simulates human thinking abilities and involves self-learning systems which leverage data mining, pattern recognition, natural language processing, text analytics, etc, similar to he way the human brain works. And this is the latest incarnation of digital banking and is based on combination of machine learning algorithms as well as rule-based solutions.

Machine learning algorithm like hidden Markov model or HMM has been a popular one in understanding sequence and has been widely used in understanding speech recognition and part-of-speech tagging needed for text analytics. The model in HMM passes through sequence of hidden states and hence the name. Refined and more nuanced applications of HMM have been reported [3, 5]. Speech recognition group at Carnegie Melon University [9] and elsewhere worked quite extensively on HMM. Next generation of machine learning algorithms are based on deep neural networks [4, 6, 7] and are generally considered to be more successful implementations of speech recognition and text analytics. For example, long short-term memory (LSTM) recurrent neural network (RNN) algorithms [11] have been used for accurate speech recognition, instead of HMM. The machine learning algorithms in these areas are widely implemented by service providers like Google Cloud Platform, Microsoft Azure, IBM Bluemix Watson, etc.

In spite of success of the algorithms for speech recognition in ideal lab environment, there have been many challenges in real implementations. The implementations require building language models with huge amount of data and building that type of dataset with accuracy for different dialects of a given language and making it speaker agnostic have been some of the biggest challenges. Because of these, many languages of the world like Hindi, Bangla [8] and many Indian languages have not been studied very well and are classified as low-resource or less studied languages. Recognizing Indian English dialects too presented significant challenges in implementation.

Millennials who love IT enabled gadgets demand new interfaces for banking and other services. Even the elderly and illiterate people can leverage the implementations as that would make life really convenient. Already video banking using mobile devices has been introduced. Here the customer interacts with a human banker through mobile app with audio-video streaming. This certainly is a great addition, but requires the presence of an efficient human banker to be present in the centralized video branch to interact with the customer and respond to queries, perform transaction and recommend banking products and solutions. Also, this cannot scale the human domain expertise. Expecting all human bankers catering to video branch banking to be well versed with product recommendations might be unrealistic. If the video branch banker is transformed into an AI powered computerized system, there would be no such limitation in scaling the efficiency of an expert banker.

As for automated recommendation system for banking products and solutions, it may be pointed out that there are two basic design approaches for a computerized recommendation system based on statistical modeling or machine learning algorithms. Content-Filtering [12] systems focus on common attributes of items and users in vector space models. Similarity of items is determined by measuring the similarity in their properties. Collaborative-Filtering [13] systems focus on the relationship between users and items. Similarity of items is determined by the similarity of the ratings of those items by the users who have rated both items. Many systems use combination of content and collaborative filtering approaches. In content filtering, a common set of terms that describe items and users is determined first. This set of terms acts the schema of content-filter based recommendation. Information about each item or user, e.g., attributes, tags, etc may be used to form a logical document representing it. If a term occurs in the logical document representing an item, its value in the vector is non-zero. Several different ways of computing these values, also known as term weights, have been developed. One of the best known schemes is term frequency inverse document frequency or TF-IDF [15] weighting. These approaches have been heavily used by e-commerce systems like Amazon, etc [14] and may be leveraged by the banking system as well for product recommendation.

Solutions of many other business problems in the bank can be mapped to use of machine learning algorithms. For example, the advanced authentication of banking user without usage of password may be done by analyzing the behavioral data using statistical approaches used in machine learning [16].

Although there are many libraries and frameworks available in different programming languages, ope-source Apache Spark has machine learning framework [17] which is very widely used. Since learning curve is stiff, practitioners often make use of the machine learning algorithms made available as services or API by third parties like Amazon Web Services or AWS, Google Cloud Platform or GCP, Microsoft AZURE, IBM Watson, etc [18]. Even robot specializing in motor or limb movement functions can be augmented with cognitive features with the help of third-party AI services [19].

Banks as pioneers in adoption of information technologies have planned solutions using machine learning algorithms, especially those related to speech recognition, natural language processing, product recommendation, etc and some have indeed rolled out first phase of implementations.

Besides satisfying the expectation of a segment of customers and improving the experience through introduction of conversational user interface, the banks are also attempting to optimize the processes and this could impact future deployment of human and non-human resources and the chief information officers of the bank would be under pressure to walk the path of optimization which might impact jobs.

In this paper, besides a review of related technologies, we focus on the technical challenges in implementation of the conversational user interface with text analytics and speech recognition in a banking ecosystem using a prototype built by us, and  share analysis of user interface based on machine learning, and also speculate on future directions.

## 2.  The Setup and Design Approach

Android mobile app was developed that captured the voice of and text entered by, the customer. The data so entered was analyzed with the help of speech recognition and text analytics services hosted on cloud service provider, Google Cloud Platform or GCP. The interaction protocol between the mobile app and the service is HTTPS.

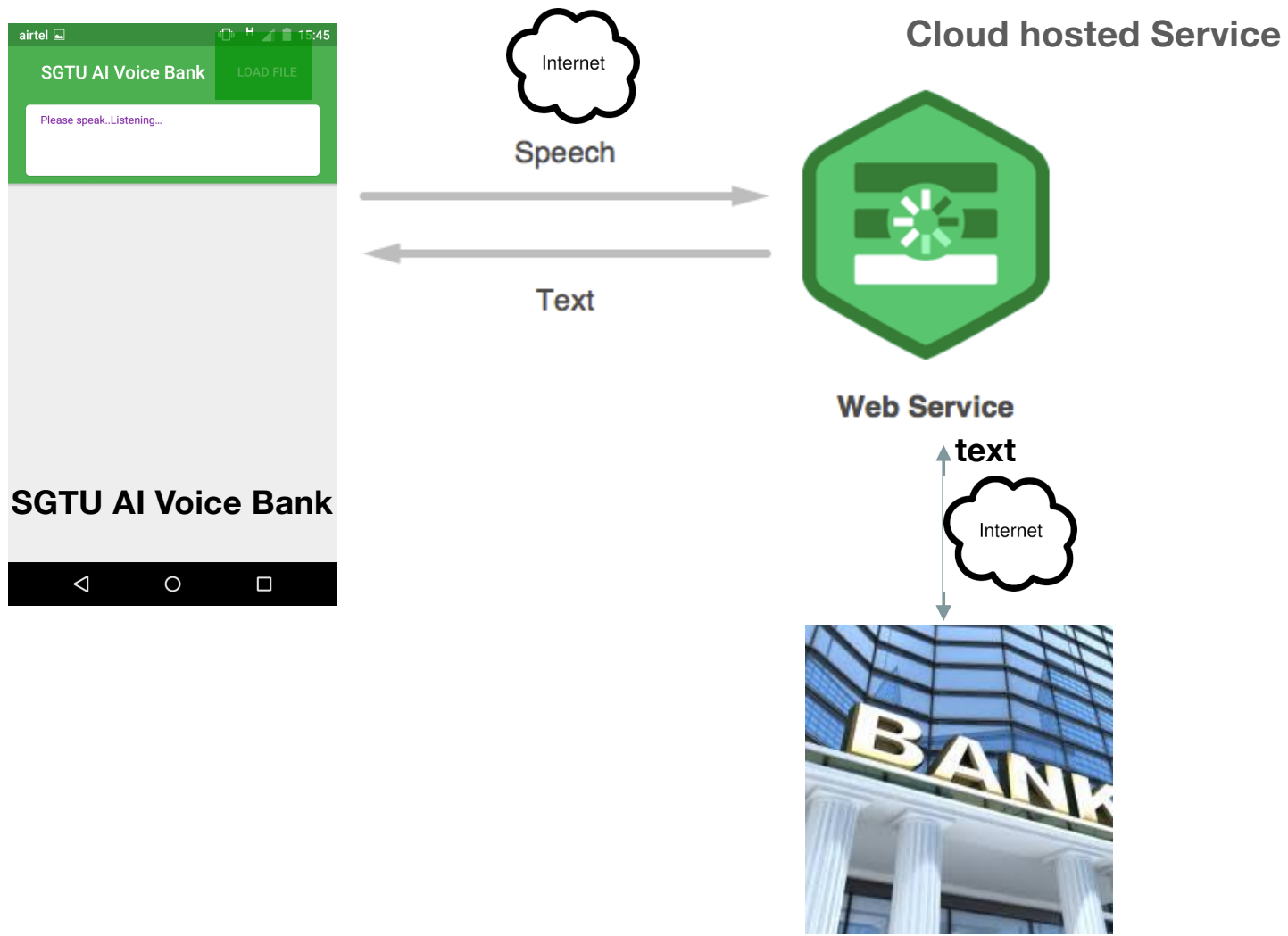

**Figure 1:** The mobile app interacting with backend speech recognition web service hosted on Google Cloud Platform (GCP). The service in turn is integrated with a simulated banking web service.

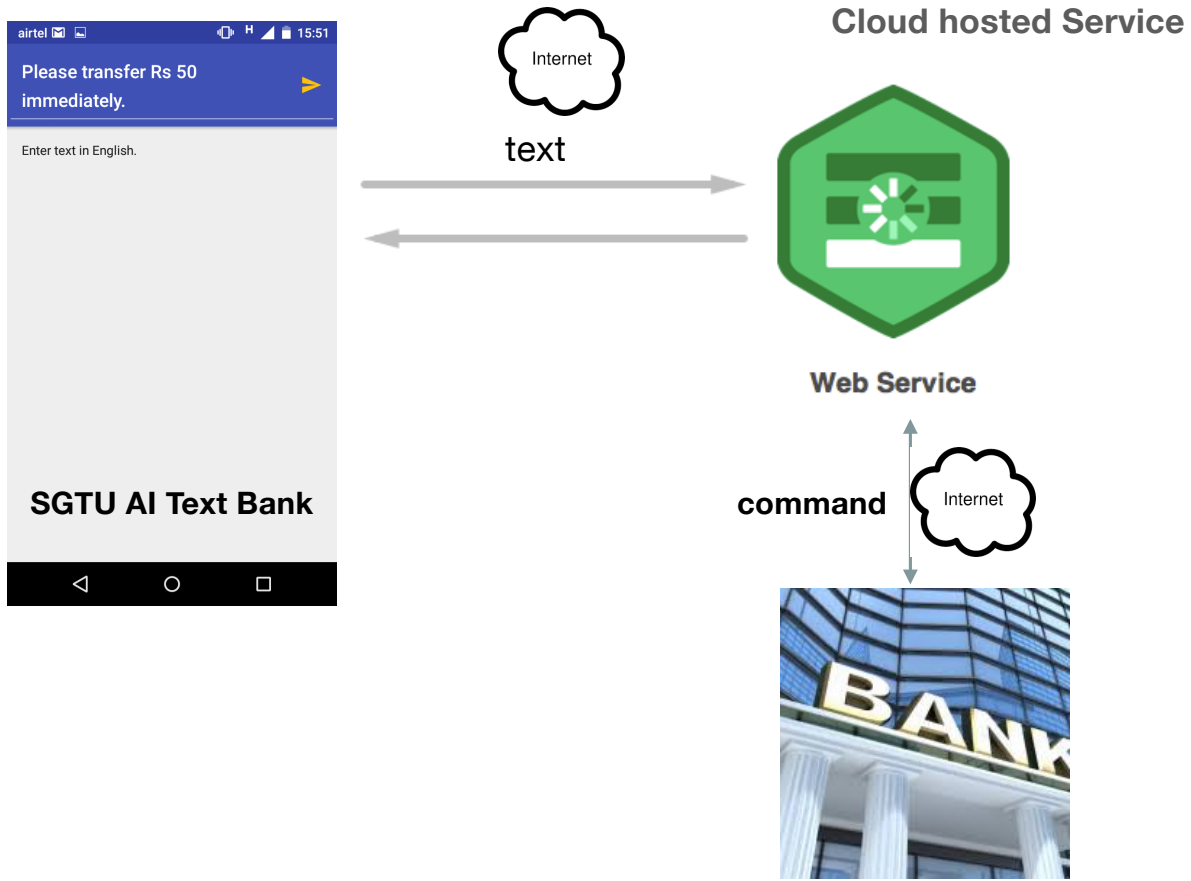

**Figure 2:** The mobile app interacting with backend text analytics web service hosted on Google Cloud Platform. The service in turn is integrated with a simulated banking web service.

The user could speak or enter text for performing simulated banking transaction and get notification of status of the transaction through registered email. The overall design is based on a mix of rule-based and cognitive programming paradigms mentioned earlier in the paper.

The setup focusses on implementation of the conversational user interface with text analytics and speech recognition in a banking ecosystem. Therefore, actual integration with a banking system was not done. Instead a RESTful web service simulating the banking system was used.

## 3.  Results and Discussion

The text analytics service worked very accurately in the sense that it is able to perform entity analysis, sentiment analysis and syntax (grammar) analysis accurately for English input and hence rule based analysis code written by us for the prototype, leveraging the results of text analytics done by the service on the GCP, was quite accurate, for almost all cases related to payment transactions . But design of a more comprehensive banking system capable of supporting a wide palette of transaction types, would involve a supervised learning algorithms which would use huge amount of training data provided by banking domain experts. For development of such a comprehensive and robust solution, the banks would have the choice of do-it-yourself (DIY) or buy a solution from a service provider like IBM Watson, Microsoft Azure, etc or their system integrator partners.

The voice recognition was less accurate for Indian accent. That is established statistically, by testing with various voice inputs from the same person at different times and also voices of different persons in the SGT University lab in Gurgaon, India. The quality of implementation of voice recognition service provided by the GCP is at par with similar offerings by competitors. The service, of course, works as a blackbox from the perspective of the consuming applications like our prototype, and the consuming application has no way of knowing the detailed algorithm employed by the service provider like GCP. However, it is fair to say that it is employing the latest and greatest state of the art in this domain and would most probably involve a combination of HMMs and long short-term memory (LSTM) recurrent neural network (RNN) algorithms [11]. In spite of that, the reliability of the system for major Indian languages like Hindi and Indian English is still not very high, but it is fair to say that failure rate is tolerable. This only validates that many Indian languages and dialects are still low-resource areas to be developed further by the language researchers working in the domain.

In our prototype, we did not focus on speaker recognition as part of authentication because the latter is a separate focus area being pursued by information security professionals [21-25] and this area is more established than speech recognition. Also, for the same reason, other forms or authentication were not studied in our prototype.

Going beyond our prototype, we can easily say that solutions of banking product recommendation would be quite mature because of huge amount of work done in e-commerce domain. Also, the backend historical and streaming analytics have matured to a great extent in a decade and these solutions [18] are ready to be used by the banks.

Overall, because of reliability and maturity levels achieved so far, some banks in the developed countries already rolled out first phase of conversational user interface and backend analytics employing machine learning algorithms, although there is scope for further improvement. In India, the solution is still in planning stage — some big banks and also innovative small players announced plan or intention for supporting such initiatives in near future. Technological hurdles mentioned earlier would perhaps delay reliable implementation by two to five years, but the emergence of such solutions are certain.

## 4. Conclusions and Future Directions

Banks have always shown boldness in adopting cutting-edge technologies in their solutions. The recent interest of the banks in adopting conversational user interface and machine learning and analytics based automated, efficient processing started quietly, in developed countries and also in developing countries like India. However it is feared that technological hurdles would perhaps delay reliable implementation by two to five years. But it is certain that emergence of such solutions cannot be stopped. One reason for delay in successful implementation in developing countries is technological, i.e., the languages are still low-resource or less developed and the gap would be bridged to a large extent in the timeframe mentioned here. The other reasons for delay in successful implementation in developing countries are cost-benefit considerations. With the evolution of customer expectations and globalized nature of technological progress, developing countries like India would not be far behind. Adoption of such solutions would lead to improvement in customer satisfaction, redefine customer expectations and surely result in optimization of human and non-human resources. Would that also impact employment of people in the bank? That certainly cannot be ruled out.